\documentclass{optica-article}

\journal{opticajournal} 

\articletype{Research Article}

\usepackage{lineno}
\usepackage{makecell}

\begin{document}

\title{Learning high-dimensional quantum entanglement through physics-guided neural networks}

\author{Yang Xu,\authormark{1,$^\dagger$, *} Hao Zhang,\authormark{2,3,$^\dagger$}  Wenwen Zhang,\authormark{2} Luchang Niu,\authormark{1} Mahtab Amooei,\authormark{4} Sergio Carbajo,\authormark{2,3,5}  and Robert W. Boyd \authormark{1,4,6}}

\address{\authormark{1}Department of Physics and Astronomy, University of Rochester, Rochester, New York 14627, USA\\
\authormark{2}Department of Electrical and Computer Engineering, UCLA, Los Angeles, California 90095, USA\\
\authormark{3}SLAC National Accelerator Laboratory, Menlo Park, California 94025, USA\\
\authormark{4}Department of Physics, University of Ottawa, Ottawa, Ontario K1N 6N5, Canada\\
\authormark{5}Department of Physics and Astronomy, UCLA, Los Angeles, California 90095, USA\\
\authormark{6}The Institute of Optics, University of Rochester, Rochester, New York 14627, USA\\
\authormark{$\dagger$}These authors contributed equally.}

\email{\authormark{*}yxu100@ur.rochester.edu} 


\begin{abstract*} 
High-gain spontaneous parametric down-conversion (SPDC) produces bright squeezed vacuum with rich high-dimensional entanglement, but its output is inherently multimodal and non-perturbative, making full (radial $\times$ azimuthal) modal characterization a major computational bottleneck. We propose a physics-guided deep neural network that reconstructs the source’s \emph{modal fingerprint}: the high-dimensional correlation signature across radial and azimuthal indices. We designed a FiLM-modulated convolutional architecture that predicts the joint \((m,l)\) distribution, and training is driven by a hybrid loss that couples data-driven metrics (JSD, KL, MSE, Wasserstein) with a soft orbital-angular-momentum (OAM) conservation term, providing an essential inductive bias toward physically consistent solutions. Across gain regimes, our method achieves high-fidelity reconstruction with average JSD of \(1.96\times10^{-3}\), WEMD of \(1.54\times10^{-3}\), and KL divergence of \(7.85\times10^{-3}\), delivering an approximate \(128\times\) speedup over full numerical simulation and more than \(30\%\) accuracy gains over U-Net baselines. These results demonstrate that physics-guided learning, via a soft OAM-conservation regularizer and physically generated training targets, enables rapid and data-efficient modal characterization. Compared with traditional numerical simulation, our \emph{mesh-free} method has demonstrated good generalization with limited or contaminated training data and has enabled fast “online” prediction of the quantum dynamics of a high-dimensional entanglement system for real-world experimental implementation. 

\end{abstract*}

\section{Introduction}\label{sec1}

Spontaneous parametric down-conversion (SPDC) has become a standard method for generating photon pairs entangled in various degrees of freedom such as time-frequency \cite{PhysRevA.47.R2472, PhysRevA.54.R1}, position-momentum \cite{PhysRevLett.92.127903, PhysRevA.76.062305}, polarization \cite{PhysRevA.50.5122,PhysRevA.69.041801}, and angle-orbital angular momentum (OAM) \cite{ mair2001entanglement,fickler2012quantum, romero2012increasing, spec}. During the past two decades, much interest has been attracted to bright squeezed vacuum (BSV), a nonclassical state of light with a macroscopic photon number that can be produced by SPDC in the high gain regime. Consequently, a plethora of experimental and theoretical works have also shown the potential of BSV in quantum sensing \cite{lawrie2019quantum}, quantum metrology \cite{brida2010detection}, quantum imaging \cite{brida2010experimental, lopaeva2013experimental}, and quantum-state engineering \cite{hurvitz2023frequency}. In addition, the large photon-number correlation in BSV leads to richer phenomena than the two-photon entanglement produced by low-gain SPDC \cite{sharapova2015schmidt}. 

Recently, the brightness of the BSV state has opened exciting new venues in ultrafast nonlinear optics and strong-field quantum optics \cite{PhysRevLett.102.183602, PhysRevA.82.011801}. This major development of an intense, non-classical light source has completely altered the understanding of many nonlinear optical processes such as high-harmonic generation (HHG) \cite{rasputnyi2024high, lemieux2025photon} and above-threshold ionization (ATI) \cite{lyu2025effect} driven by intense non-classical light in various materials. On a more fundamental level, the nonclassical state of light, such as BSV, can introduce additional modulations to the electron trajectories different from those driven by classical coherent laser pulses \cite{heimerl2025quantum}. In this strong driving field regime, a full theoretical description of the quantum state of the entangled photon pairs in BSV remains complicated because the standard perturbation treatment breaks down \cite{sharapova2020properties}. Furthermore, since the generation of BSV is often achieved by pumping the SPDC crystal with intense ultrashort pulses, the generated BSV pulse is inherently in a multimodal squeezed state. It usually consists of multiple squeezed modes in one pulse, each of which has its own degree of squeezing. Hence, although an efficient mode characterization of high-gain SPDC sources is a critical prerequisite for most squeezing-based experiments, such as sub-shot-noise measurement \cite{xiao1987precision, pooser2015ultrasensitive, oelker2016ultra} and mode correlation measurement \cite{cutipa2022bright}, the non-perturbative nature \cite{even2024motion} and the high-dimensional multimodal squeezing have made the full mode characterization of high-gain SPDC a daunting computational and experimental task \cite{PhysRevA.83.033816, PhysRevA.106.063709}. Therefore, developing efficient techniques for multimodal characterization is required.

Over the past decade, deep learning has become a powerful tool for handling high-dimensional, nonlinear problems in optics-related fields like computational imaging~\cite{barbastathis2019use,zhang2025hybrid,lin2018all}, optical sensing~\cite{zhang2025accurate,yuan2023geometric,zuo2022deep}, and quantum photonics~\cite{kudyshev2020machine,ji2023recent}. It is especially useful when traditional models fall short—either because of the complexity of the system or the intricacy of the governing equations~\cite{genty2021machine,salmela2021predicting}. In optics and quantum information science, deep neural networks are now being used alongside with, or even instead of, traditional techniques like quantum state tomography~\cite{schmale2022efficient,quek2021adaptive}, entanglement detection~\cite{urena2024entanglement,deng2017quantum}, and system identification~\cite{youssry2024experimental,cong2019quantum}.

Generally, data-driven deep learning models often need a massive amount of data, and they do not always reflect the physical rules of the system~\cite{saba2022physics,ji2023recent,zhang2025multi}. This is a problem in both classical physics and quantum mechanics, where data are expensive to collect and every step is strictly governed by physical laws~\cite{tang2022physics}. Consequently, physics-guided neural networks (PGNNs) have been proposed to incorporate known physical laws directly into the model as the boundary~\cite{momeni2025training,ji2023recent}. Instead of learning purely from data, our model incorporates physics-informed constraints, such as spatial symmetries, conservation rules, and system-specific structure, to guide training. This approach reduces overfitting, enhances generalization with limited data, and ensures that predictions remain consistent with known physical principles. In practice, many quantum optical systems, such as entangled photon sources generated via SPDC, involve high-dimensional continuous variables (e.g., spatial mode indices) and exhibit fundamentally probabilistic behavior due to quantum uncertainty and measurement noise. These factors make accurate reconstruction of the underlying quantum states particularly challenging~\cite{chen2020physics}. For example, full quantum state tomography often requires time-consuming measurements or is computationally expensive. Deep learning models such as CNNs~\cite{lecun2010convolutional,zhang2025multi}, UNets~\cite{isensee2021nnu,zhang2025hybrid}, and autoencoders~\cite{chen2023photonic,boussafa2025deep} have shown promise by learning from fewer measurements. However, if these models are not guided by physics, their predictions can deviate from what is physically possible.

In our work, we combine deep learning with physics laws by building a physics-guided neural network (PGNN) called \textit{OAMNet}, which is designed for computationally efficient modal characterization of high-dimensional entangled states. Our physics-guided network achieves an average Jensen-Shannon divergence (JSD) of $1.96\times10^{-3}$, a Wasserstein earth mover’s distance (WEMD) of $1.54\times10^{-3}$, a Kullback-Leibler (KL) divergence of $7.85\times10^{-3}$, and an 128-fold speedup over the full numerical simulation. Our method outperforms baseline U-Net models by more than 30\% in reconstruction accuracy, which enables rapid and high-fidelity characterization of high-dimensional entangled states.

\begin{figure*}[htbp] 
\centering 
\includegraphics[width=0.9\textwidth]{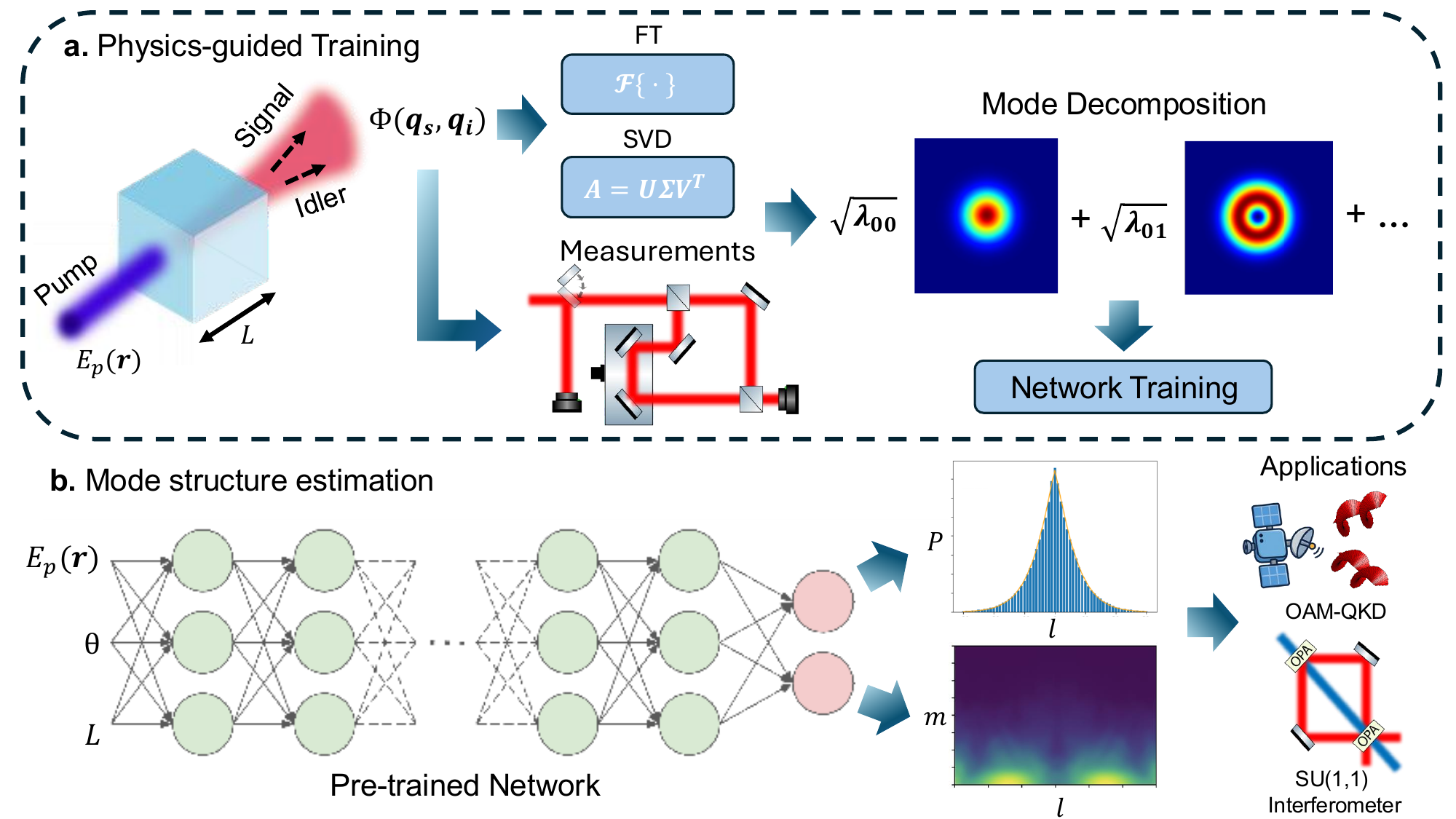} 
\caption{\textbf{Physics-guided deep learning neural network (PGNN) architecture for spatial mode distribution prediction.} \textbf{a.} Concepts of physics-guided training for SPDC modal structure estimation. In a degenerate type-I SPDC process, a crystal with a second-order nonlinearity, $\chi^{(2)}$, is pumped by a short-wavelength laser beam with a given spatial profile, $E_p(\mathbf{r})$. A signal-idler pair is generated at a lower optical frequency and its spatial mode distribution is governed by the two-photon wavefunction $\Phi(\mathbf{q}_s,\mathbf{q}_i)$. The joint spatial mode distribution, or the spatial mode structure, can be obtained through Schmidt decomposition of the two-photon wavefunction or through direct experimental measurement. The simulation and measurement results, together with soft physical regularizers (e.g. OAM conservation), are fed to the neural network during the training stage. \textbf{b.} Estimation of the mode structure using a trained PGNN. The PGNN, once properly trained with physical constraints, takes in a set of physical parameters (e.g. pump intensity profile, crystal orientation, crystal length, etc.) as an input vector and infers the full modal structure of an arbitrary SPDC setup. This computationally efficient process enables real-time feedback in applications in OAM-based QKD or SU(1,1) interferometry.}

\label{fig1} 
\end{figure*}

\section{Design of physics-guided deep neural network}\label{sec2}

\begin{figure*}[htbp] 
\centering 
\includegraphics[width=1\textwidth]{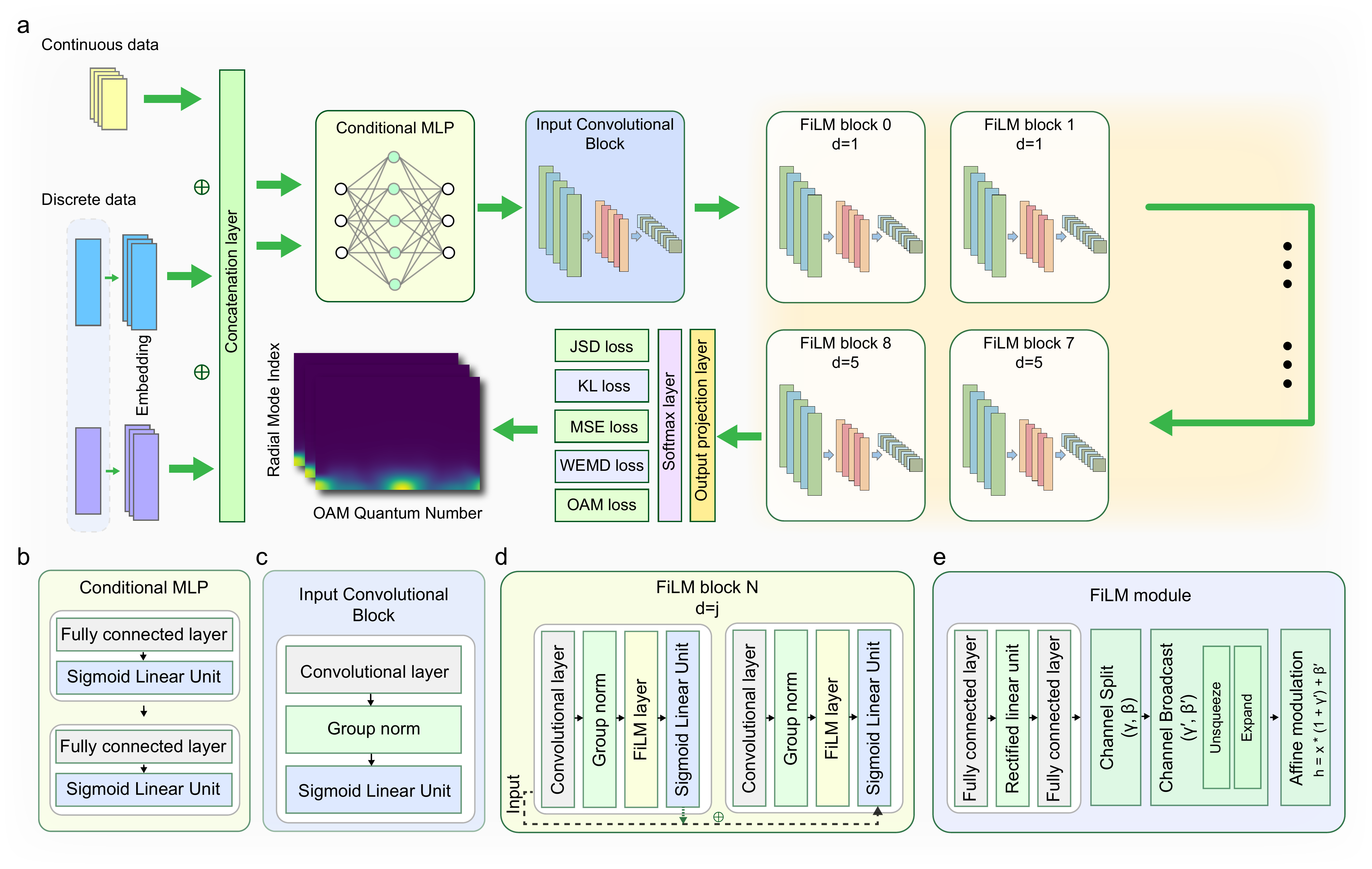} 
\caption{\textbf{Physics-guided deep learning architecture for OAM mode prediction.} \textbf{a.} Continuous parameters and embedded discrete OAM indices are concatenated to form a conditioning vector that modulates a stack of dilated convolutional FiLM residual blocks. The network outputs a normalized distribution over the $(m, l)$ grid via a final $1\times1$ convolution and softmax.  \textbf{b.} The Conditional MLP transforms the conditioning vector through two SiLU-activated fully connected layers, generating the modulation features used throughout the network. \textbf{c.} The input convolutional block applies a convolution, GroupNorm, and SiLU activation to extract the initial spatial features. \textbf{d.} A series of FiLM residual blocks with increasing dilation factors captures spatial structure across multiple scales. \textbf{e.} Each FiLM module computes affine modulation parameters from the conditioning features, allowing the convolutional backbone to adapt dynamically to the underlying physical conditions.}

\label{fig2} 
\end{figure*} 

For a type-I SPDC process, the Hamiltonian in the interaction picture is 
\begin{align}
        \mathcal{H} = i\hbar\Gamma \int d^3\mathbf{q}_i d^3\mathbf{q}_s \Phi(\mathbf{q}_s, \mathbf{q}_i)  \hat{a}_{\mathbf{q}_s}^\dagger \hat{a}_{\mathbf{q}_i}^\dagger +\text{h.c.} \label{eq:hamilt}
\end{align}
where $\Gamma$ is the nonlinear coupling constant, $\mathbf{q}_{s,i}$ is the transverse wavevector of the generated signal/idler photon and $\hat{a}^{\dagger}_{\mathbf{q}_{s,i}}$ is the creation operator of a signal/idler photon with a transverse wavevector $\mathbf{q}_{s,i}$. $\Phi$ is the two-photon wavefunction describing the quantum state of the photon pair generated from SPDC
\begin{align}
       \Phi(\mathbf{q}_s, \mathbf{q}_i) = C \tilde{E_p}(\mathbf{q}_s + \mathbf{q}_i) \text{sinc}\left(\frac{\Delta k L}{2}\right)e^{\frac{i\Delta k L}{2}}\label{eq:two-photon-wavefunction}
\end{align}
where $C$ is the normalization constant, $L$ is the length of the nonlinear crystal and $\Delta k = k_{pz} - k_{sz} - k_{iz}$ describes the phase mismatch.

From the two-photon wavefunction $\Phi(\mathbf{q}_s, \mathbf{q}_i)$, the full spatial mode distribution of the entangled photon pairs can be determined by Schmidt decomposing $\Phi(\mathbf{q}_s, \mathbf{q}_i)$ in a complete set of orthonormal spatial mode basis (Supplemental 1):
\begin{align}
        \Phi(\mathbf{q}_s, \mathbf{q}_i) = \sum_{ml} \tilde{\lambda}_{ml} u_{ml}(q_s) v_{ml}(q_i)e^{il(\phi_s - \phi_i)}\label{eq:decompose}
\end{align}
where $\tilde{\lambda}_{ml}$ is the normalized eigenvalue, which shows the weight of each mode $(m,l)$, and $u_{ml}(q_s)$ $(v_{ml}(q_i))$ is the radial eigenmode of the signal (idler). $m$ is the radial mode index, and $l$ stands for the azimuthal mode index, often known as the OAM number. The spatial mode distribution is usually conveniently characterized by the effective mode number, also known as the Schmidt number, $K$. The Schmidt number is a critical metric to measure the effective dimensionality of high-dimensional entanglement in quantum communication \cite{miatto2012spatial} and is defined as  
\begin{align}
    K = \frac{1}{\sum_{ml} \tilde{\lambda}_{ml}^2}
\end{align} where the weight $\tilde{\lambda}_{ml}$ is normalized such that $\sum_{ml}\tilde{\lambda}_{ml} = 1$. The computation of the full modal structure and the Schmidt number of high-dimensional entanglement state involves multi-dimensional numerical integration and singular value decomposition of large matrices, thus leading to extremely high time complexity (see Methods).

As a feasible solution, we employ a deep learning neural network based on a Feature-wise Linear Modulation (FiLM)-conditioned residual architecture, as shown in Figure~\ref{fig2}a. The model takes two types of input parameters: continuous parameters (e.g. pump power and crystal orientation) and discrete variables (e.g. spatial mode index of the pump). The output is a normalized 2D probability distribution representing the complete Schmidt mode structure of down-converted photons, parameterized over radial ($m$) and orbital angular momentum ($l$) modes.

The architecture begins with embedding layers for the discrete OAM indices (crystal and pump), concatenated with the standardized continuous parameters to form the conditioning vector $\mathbf{c}$. To prepare $\mathbf{c}$ for modulation of the convolutional backbone, Figure~\ref{fig2}b shows that it is first processed by a two-layer Conditional MLP with SiLU activations, producing a compact nonlinear representation of the experimental parameters for FiLM modulation. These features are then passed to the Input Convolutional Block in Figure~\ref{fig2}c, which applies a convolution, GroupNorm, and SiLU activation to map the $(m, l)$ grid into a feature space suitable for the subsequent FiLM-modulated residual layers.

The core model is a ResNet-style variant, in which each residual block is modulated by FiLM layers, as shown in Figure~\ref{fig2}d and e. The FiLM mechanism dynamically adjusts intermediate feature maps based on the conditioning input by applying affine transformations to each channel. Given a condition vector $\mathbf{c}$ derived from the input parameters, the FiLM layer computes scale ($\gamma$) and shift ($\beta$) parameters through a small MLP:
\begin{equation}
\gamma, \beta = \text{MLP}(\mathbf{c})
\end{equation}
These are applied to intermediate features $\mathbf{h}$ by the rule:
\begin{equation}
\text{FiLM}(\mathbf{h}; \gamma, \beta) = \mathbf{h} \cdot (1 + \gamma) + \beta
\end{equation}

This vector modulates a stack of convolutional FiLM residual blocks with increasing dilation rates, allowing the network to capture both local and global spatial dependencies. The final prediction is produced by a $1\times1$ convolution followed by a softmax operation across the $(m, l)$ grid, ensuring a normalized output. Considering the output should represent a physically constrained joint distribution, purely data-driven training may be insufficient: the network may fit the data but have the possibility to violate fundamental physical laws. To prevent such physically inconsistent solutions, we adopt a physics-guided training scheme that embeds domain constraints, the conservation of OAM \cite{kopf2025conservation}, into the loss function. By combining losses with a soft OAM-conservation penalty, the model is encouraged to learn solutions that are not only statistically accurate but also physically permissible in the real world. In short, the network functions as a physics-guided surrogate model trained on simulation/experimental data generated from the SPDC Hamiltonian and phase-matching physics. The primary explicit constraint incorporated during training is a soft OAM-conservation loss, rather than complicated enforcement of the full Hamiltonian dynamics within the network architecture.

\section{Implementation of OAMNet}\label{sec3}
\subsection{Results}
\begin{figure*}[htbp] 
\centering 
\includegraphics[width=1\textwidth]{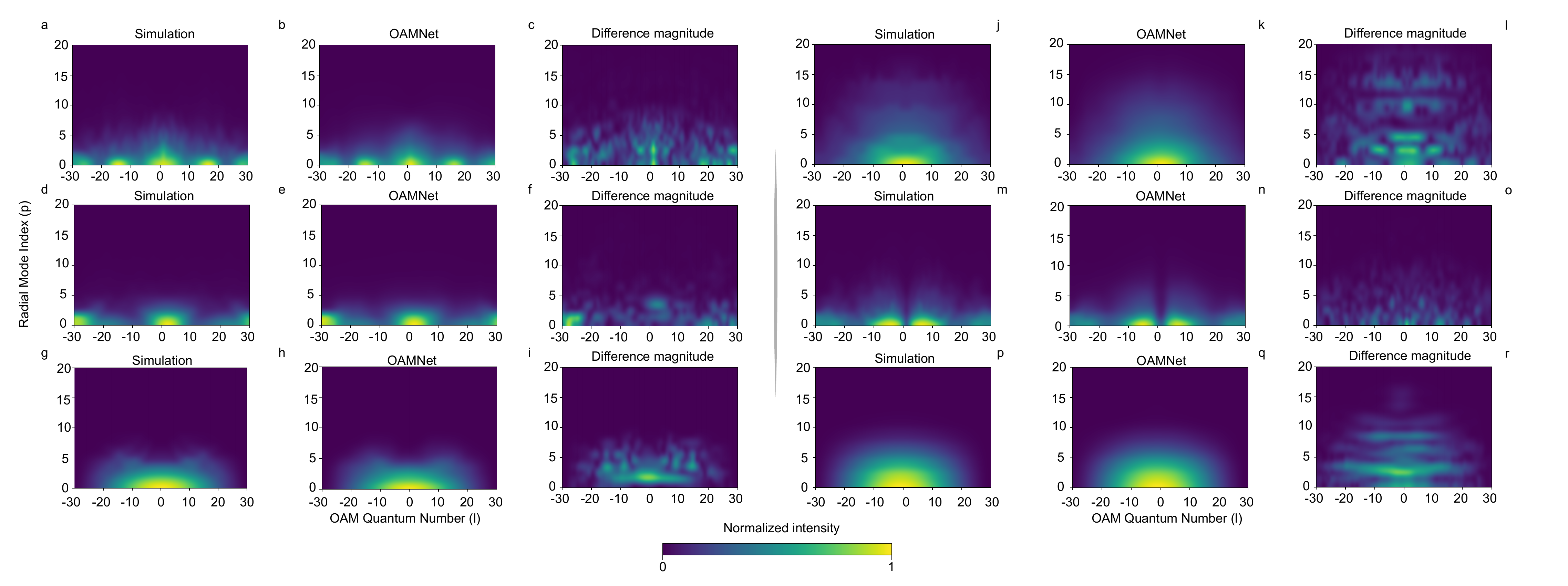} 
\caption{\textbf{Comparison between simulated OAM mode distributions and OAMNet predictions.}
(a--r) Representative examples showing the agreement between simulated $(m,\ell)$ intensity distributions (left of each pair) and the corresponding OAMNet predictions (right). Color denotes normalized intensity.
} 
\label{fig3} 
\end{figure*}

Figures~\ref{fig3}a–r compare the randomly selected examples of reconstructed joint probability distributions of the two-photon spatial modes in the LG basis.
Each row corresponds to a distinct set of physical and pump parameters, shown in Table ~\ref{tab:0}.

\begin{table}[ht]
\centering
\caption{Physical parameters for the reconstructed joint probability distributions shown in Fig.~\ref{fig3}.}
\begin{tabular}{ccccccc}
\hline
\textbf{Subfigures} & $g$ & $\theta\ (^\circ)$ & $L$ & $w_p$ & $\ell_p$ & $p_p$ \\
\hline
a--c & 0.021 & 32.929 & 3196.603 & 142.043 & 2  & 4 \\
d--f & 0.364 & 32.985 & 3494.908 & 138.649 & 4  & 4 \\
g--i & 1.501 & 32.925 & 890.232  & 667.724 & 0  & 1 \\
j--l & 5.364 & 32.951 & 997.411  & 358.182 & 2  & 3 \\
m--o & 0.510 & 32.923 & 2596.429 & 134.738 & 2  & 3 \\
p--r & 0.130 & 32.990 & 2316.258 & 299.028 & -2 & 0 \\
\hline
\end{tabular}
\label{tab:0}
\end{table}

Here, $g$, $w_p$, $l_p$ and $p_p$ denote the parametric gain, beam waist, the OAM mode index and the radial mode index of the pump. $L$ and $\theta$ are the crystal length and the angle between the pump propagation direction and the crystal optical axis, respectively. 
For each representative configuration, the simulation results are juxtaposed with the corresponding predictions by OAMNet. The vertical axis denotes the radial mode index $m$, while the horizontal axis represents the OAM quantum number $l$. The intensity indicates the normalized joint probability amplitude. Across all tested scenarios, OAMNet successfully captures the intricate modal features of the simulated distributions, including the localization of the dominant modes in both radial and azimuthal dimensions. Even in cases involving sharply peaked or asymmetric mode structures (e.g., Figures~\ref{fig3} c–d and e–f), the predictions exhibit high fidelity to the ground truth, preserving both the spatial support and shape of the entangled state distributions.
To further benchmark the performance, we perform quantitative comparisons between three model configurations: a lightweight UNet architecture, a deeper unconstrained UNet, and our proposed physics-guided network, OAMNet.

\begin{figure*}[htbp] 
\centering 
\includegraphics[width=1\textwidth]{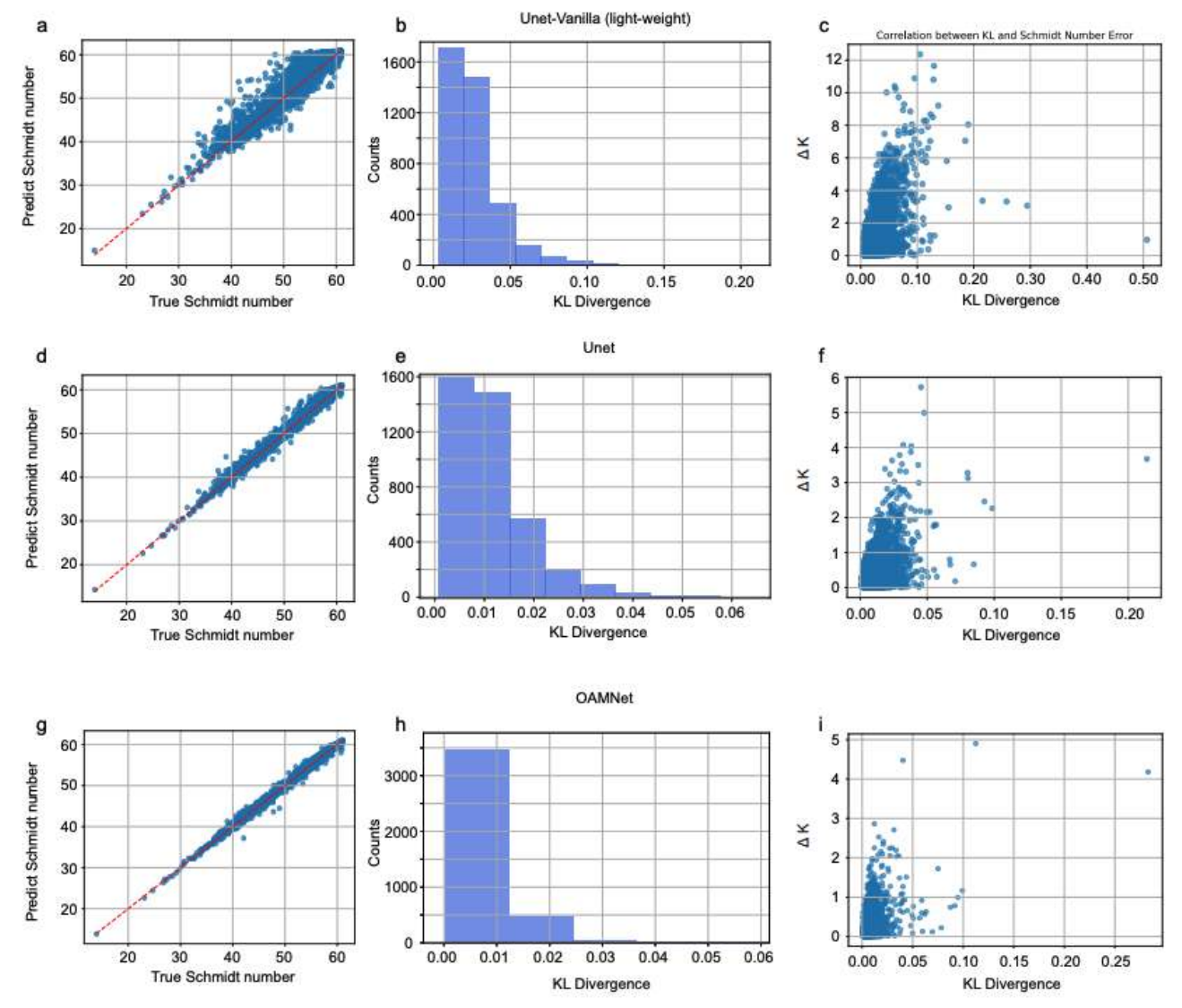} 
\caption{\textbf{Comparison of U-Net Vanilla (light-weight), U-Net, and OAMNet on Schmidt number prediction and KL divergence statistics.}\textbf{a--c}. \textbf{U-Net Vanilla (light-weight):} \textbf{a}. Predicted versus true Schmidt numbers; \textbf{b}. Histogram of KL divergences between predicted and simulated $(m,\ell)$ distributions; \textbf{c}. Scatter plot of Schmidt number error $\Delta S$ versus KL divergence. \textbf{d--f}. \textbf{U-Net:} Corresponding plots showing \textbf{d}. predicted versus true Schmidt numbers, \textbf{e}. KL divergence distribution, and \textbf{f}. $\Delta S$ versus KL divergence.
\textbf{g--i}. \textbf{OAMNet:} \textbf{g}. Predicted versus true Schmidt numbers, \textbf{h}. KL divergence histogram, and \textbf{i}. $\Delta S$ versus KL divergence.
} 
\label{fig4} 
\end{figure*}

Figure~\ref{fig4} displays the results of this comparison across three key dimensions: prediction-true of the  Schmidt number, Kullback-Leibler (KL) divergence, and the relationship between distributional error and entanglement estimation. The first row (Figure~\ref{fig4} a–c) corresponds to the Light-weight UNet, which exhibits significant scatter in its Schmidt number predictions and a relatively broad KL divergence distribution, indicating limited capacity to learn the full structure of the data. The error in the predicted Schmidt number correlates strongly with KL divergence, suggesting that poorer distribution matching leads directly to less accurate entanglement estimation.
The second row (Figure~\ref{fig4} d–f) corresponds to the deeper UNet model, which provides improved performance across all metrics. The Schmidt number predictions are more tightly clustered along the identity line, the KL divergence histogram is narrower and more peaked, and the error correlation is reduced. However, it is the final OAMNet model, shown in Figure~\ref{fig4} g–i, that achieves the highest accuracy. Predictions of the Schmidt number are the most consistent with ground truth values, the KL divergence values are sharply concentrated near zero, and the $\Delta$ Schmidt vs. KL divergence plot reveals minimal error spread. These improvements underscore the importance of incorporating physical knowledge into the learning process. By enforcing constraints such as OAM conservation and mode normalization, OAMNet not only better approximates the data distribution but also preserves critical physical observables that define the nature of high-dimensional entanglement than the general-purpose neural networks.

Table~\ref{tab:1} summarizes the computational and predictive performance of different models measured by multiple metrics. The baseline simulation, while accurate, demands significant memory (98\% RAM) and computation time (38 seconds per sample). In contrast, the lightweight Vanilla-style UNet achieves fast inference ($<$1s) and modest memory usage, but its prediction accuracy is limited (Avg JSD: $6.81 \times 10^{-3}$). A deeper standard UNet (1.18M parameters) improves performance substantially (JSD: $2.86 \times 10^{-3}$) with slightly higher GPU and RAM usage. Our physics-guided model (950K parameters) balances complexity and physical fidelity, achieving the best accuracy across all metrics (JSD: $1.96 \times 10^{-3}$, WEMD: $1.54 \times 10^{-3}$, KL: $7.85 \times 10^{-3}$) with efficient resource usage. 

\begin{table*}[]
\centering
\footnotesize
\caption{Comparison of different computing algorithms in terms of number of parameters, time complexity, spatial complexity and average performance metrics (JSD, WEMD, KL and $\Delta K$).}
\label{tab:1}

\begin{tabular}{l|l|l|l|l|l|l|l|l|l|l}
\hline
\textbf{} &
  \textbf{\makecell[l]{Param. \\ count\\ ($10^{6}$)}} &
  \textbf{\makecell[l]{Train \\ time}} &
  \textbf{\makecell[l]{Comp. \\ time}} &
  \textbf{\makecell[l]{RAM \\ usage}} &
  \textbf{\makecell[l]{GPU \\ usage }} &
  \textbf{\makecell[l]{CPU \\ usage }} &
  \textbf{\makecell[l]{Avg \\ JSD\\ ($10^{-3}$)}} &
  \textbf{\makecell[l]{Avg \\ WEMD \\($10^{-3}$)}} &
  \textbf{\makecell[l]{Avg \\ KL \\($10^{-3}$)}} &
  \textbf{\makecell[l]{Avg \\$\Delta$K}} \\
\hline
\textbf{Simulation} & - & - & 128 s & 98\% & - & 92\% & - & - & - & - \\ \hline
\textbf{U-Net Vanilla} & 0.07  & 5883 s & $<1$ s & 72\% & 84\% & 15\% & 6.81 & 3.69 & 27.2 & 0.02 \\ \hline
\textbf{U-Net} & 1.18  & 7478 s & $<1$ s & 86\% & 88\% & 12\% & 2.86 & 2.08 & 11.5 & 0.011 \\ \hline
\textbf{OAMNet} & 0.95  & 7390 s & $<1$ s & 78\% & 86\% & 14\% & 1.96 & 1.54 & 7.85 & 0.008 \\ \hline
\end{tabular}

\end{table*}

\subsection{Ablation studies}
Ablation results provide mechanistic evidence for how each loss term shapes reconstruction fidelity and physical consistency. Hence, we carried out ten controlled experiments (E0 – E9). In every run the JSD retained a fixed weight of 1.0, while the weights attached to KL divergence, MSE, WEMD, and the OAM constraint were systematically varied. Performance was assessed with four aggregate metrics: the average JSD, KL, and WEMD over the test set, together with the absolute error in the Schmidt number ($\Delta$K). Table \ref{table:ablation} presents the numerical outcomes.

Starting from the JSD-only baseline (E0), we found that the network already reproduced the coarse shape of the joint OAM spectrum. Adding a modest KL term (E1) immediately reduced both JSD and KL, demonstrating that KL is particularly effective at correcting the underestimated tails that dominate entangled states. Introducing an MSE penalty instead (E2) led to slightly smoother maps, evidenced by modest reductions in JSD and KL, though the global WEMD barely changed; the metric therefore acts mainly as a local regularizer.

When the WEMD term was activated on its own (E3), the global structure of the distribution aligned better with ground truth, but the tails remained sub-optimal. In contrast, replacing all data-driven auxiliaries with the OAM constraint alone (E4) produced the poorest data fit, confirming that physical consistency cannot substitute for statistical supervision.

The most instructive behaviour appears when the full set of data terms is combined (E5) and the OAM loss is introduced gradually (E6 to E9). A moderate physical weight of 0.05 to 0.10 (E6 and E7) tightened the average WEMD and KL by roughly ten percent while preserving an excellent Schmidt-number error ($\Delta$K $=$ 0.008). Beyond this interval, however, stronger enforcement (E8 and E9) reversed the gains and raised all data-based losses, revealing a clear trade-off between empirical accuracy and rigid physical adherence.
Overall, the configuration adopted in our main model (E7) provides the best compromise: it delivers the lowest KL and the lowest WEMD of the study, coupled with the smallest JSD among all physically informed variants, all while maintaining the target Schmidt number. These observations confirm that each auxiliary term addresses a distinct failure mode, KL sharpens rare-event statistics, MSE polishes local continuity, WEMD secures large-scale geometry, whereas the OAM prior is indispensable for respecting conservation laws but must be applied with restraint.

\begin{table*}[t]
\centering
\caption{Ablation study of different settings.}
\label{tab:2}
\makebox[\textwidth][c]{%
  \begin{tabular}{lllllllll}
    \hline
    \textbf{ID} & \textbf{JSD} & \textbf{KL} & \textbf{MSE} & \textbf{WEMD} & \textbf{\makecell[l]{OAM \\ Loss}} & \makecell[l]{Avg JSD \\ ($10^{-3}$)} & \makecell[l]{Avg WEMD \\ ($10^{-3}$)} & \makecell[l]{Avg KL \\ ($10^{-3}$)} \\
    \hline
    \textbf{E0} & 1 & 0 & 0 & 0 & 0 & 2.38 & 1.78 & 9.58 \\
    \textbf{E1} & 1 & 0.2 & 0 & 0 & 0 & 2.02 & 1.56 & 8.08 \\
    \textbf{E2} & 1 & 0 & 0.2 & 0 & 0 & 2.11 & 1.63 & 8.49 \\
    \textbf{E3} & 1 & 0 & 0 & 0.5 & 0 & 2.19 & 1.59 & 8.82 \\
    \textbf{E4} & 1 & 0 & 0 & 0 & 0.1 & 2.66 & 2.30 & 10.1 \\
    \textbf{E5} & 1 & 0.2 & 0.2 & 0.5 & 0 & 2.17 & 1.62 & 8.65 \\
    \textbf{E6} & 1 & 0.2 & 0.2 & 0.5 & 0.05 & 1.99 & 1.50 & 7.99 \\
    \textbf{E7} & \textbf{1} & \textbf{0.2} & \textbf{0.2} & \textbf{0.5} & \textbf{0.1} & \textbf{1.96} & \textbf{1.54} & \textbf{7.85} \\
    \textbf{E8} & 1 & 0.2 & 0.2 & 0.5 & 0.2 & 2.25 & 1.90 & 9.03 \\
    \textbf{E9} & 1 & 0.2 & 0.2 & 0.5 & 0.3 & 2.35 & 2.13 & 9.44 \\
    \hline
  \end{tabular}%
}
\end{table*}

\section{Experimental validation of physics-guided OAMNet}\label{sec4}
Although the complete Schmidt decomposition yields the most accurate description of the mode structure of the entangled SPDC photons, the use of OAM modes is often more feasible in real-world applications such as quantum key distribution (QKD) \cite{PhysRevLett.113.060503}. In an SPDC process driven by a pump with a Gaussian mode profile ($l=0$), the conservation of OAM leads to a final state that is a high-dimensional superposition of signal and idler modes with anti-correlated OAM values \cite{mair2001entanglement, PhysRevLett.101.163602, PhysRevLett.134.203601}. Thus, the OAM spectrum plays a critical role in the characterization of the high-dimensional entanglement produced by SPDC sources. The width of the distribution, the so-called "spiral bandwidth", has been chosen as a practical measure to describe the effective dimensionality of the Hilbert space in which a quantum communication protocol is established.

\begin{figure*}[htbp] 
\centering 
\includegraphics[width=1\textwidth]{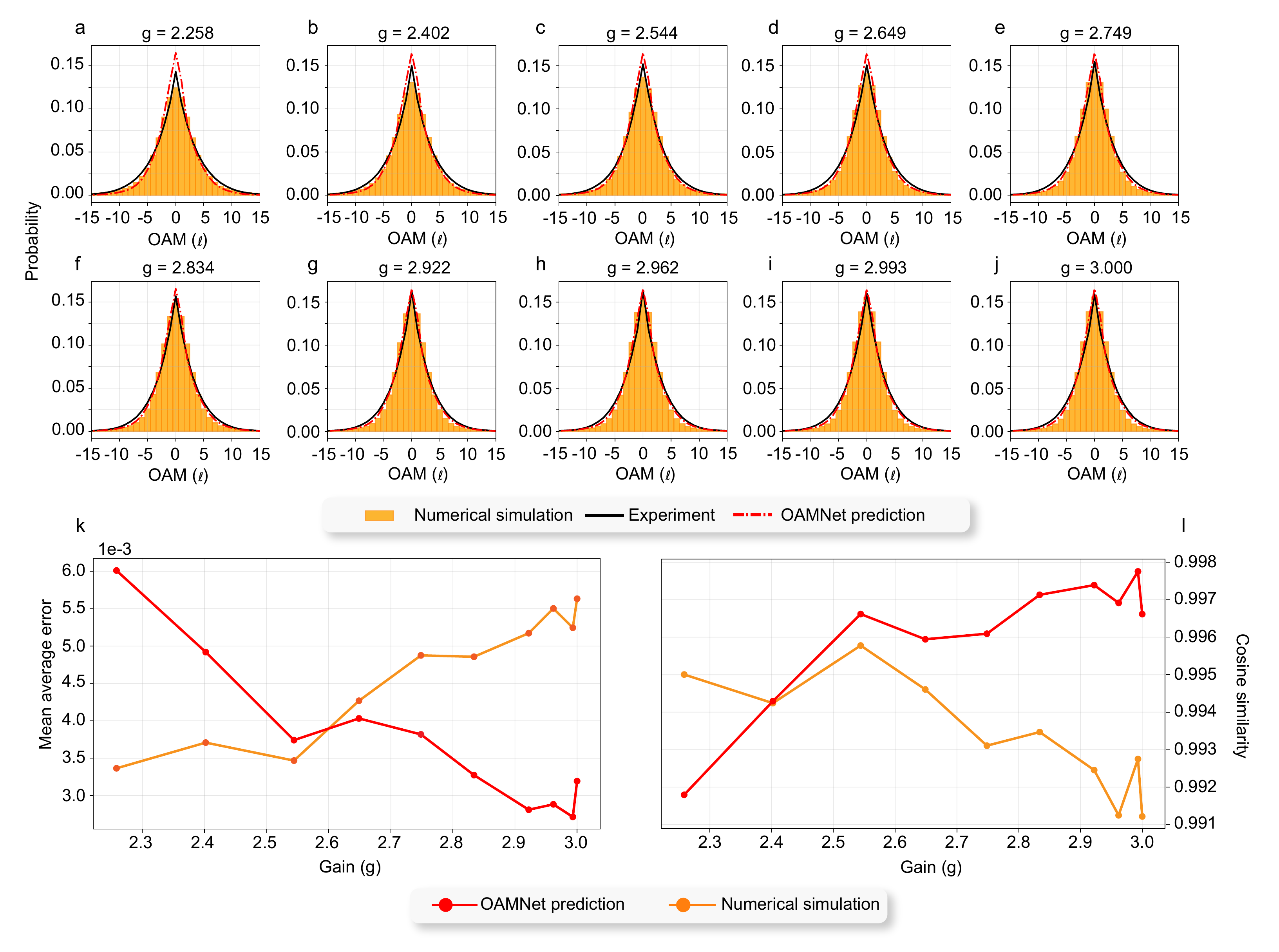} 
\caption{\textbf{Comparison of OAM spectra for varying gain values $g$.} \textbf{a–j}. show the predicted marginal OAM spectra (red dash-dotted lines), numerical simulation (orange bars), and experimental data (black solid lines) for a range of gain values $g \in [2.258, 3.000]$. 
    Excellent agreement is observed across all gains. 
    (k) plots the mean absolute error (MAE), defined as $\mathrm{MAE} = \langle |S_\ell^{\rm model} - S_\ell^{\rm exp}|\rangle$, for both OAMNet predictions and numerical simulations against experimental measurements.
    The errors remain below $6 \times 10^{-3}$ across all gains. 
    (l) shows the cosine similarity between predictions/simulations and experimental spectra: 
    $\mathrm{cos\_sim} = \frac{\sum_\ell S_\ell^{\rm model} S_\ell^{\rm exp}}{\|S^{\rm model}\|\;\|S^{\rm exp}\|}$, 
    indicating high similarity ($\ge 0.991$) throughout the range.
} 
\label{fig5} 
\end{figure*} 

The OAM spectrum $S(l)$ of a type-I collinear SPDC can be calculated from the following integral
\begin{align}
    S(l) = \left|\int_0^{2\pi}W(\Delta\phi) e^{il\Delta \phi} d\Delta\phi\right|^2
\end{align}
where $W(\Delta \phi) = \iint dq_sdq_i q_i q_s\Phi(q_s, q_i, \Delta\phi)$. $\Phi(q_s, q_i, \Delta\phi)$ is the two-photon wavefunction, Eq. \ref{eq:two-photon-wavefunction}, in cylindrical coordinates. Here, rotational symmetry is assumed so that $\Phi(q_s, q_i, \Delta\phi)$ depends only on the difference between the azimuthal angle of the signal and idler wavevectors, namely, $\Phi(q_s, q_i, \phi_s, \phi_i) = \Phi(q_s, q_i, \Delta\phi)$ where $\Delta\phi = \phi_s -\phi_i$ (Supplement 1).

With the ability to accurately estimate the complete spatial mode distribution of the SPDC two-photon wavefunction $\Phi$, our PGNN can predict the OAM spectrum with minimal additional computational cost. For each OAM mode $l$, we obtain its probability density $S(l)$ by summing the contributions of each radial mode $m$. The procedure is equivalent to calculating the marginal probability distribution of $l$, $S(l) = \sum_m \tilde{\lambda}_{ml}$. 

As a part of our experimental validation, we employ the interferometric technique \cite{kulkarni2017single} to measure the OAM spectrum of high-gain SPDC field. As shown in the experimental setup in Fig. S1 in Supplement 1, we interfere the far-field SPDC signal with its flipped mirror image on an EMCCD camera. We then take the difference image between the constructive and destructive interferograms to reconstruct the angular coherence function, from which the OAM spectrum can be retrieved by taking the Fourier transform of the azimuthal coordinate (see details in Methods). The experiment is performed on a type-I BBO with collinear phase-matching geometry and is repeated for different pump amplitudes. In Fig.~\ref{fig5}, we compare our PGNN prediction of the OAM spectrum for various parametric gains $g$ with experimental measurements and calculations based on quantum theory. For parametric gains $g$ ranging from 2.26 to 2.75, we observe good alignment between our neural network predictions and the measured OAM spectra. Fig.~\ref{fig5}k reports the mean absolute error (MAE) between the predicted or simulated distributions and the experimental measurements. The MAE remains on the order of $10^{-3}$ for all cases, demonstrating that the PGNN model captures the key features of the two-photon wavefunction across a broad range of physical parameters. Additionally, the cosine similarity in Fig.~\ref{fig5}l further supports the high agreement between prediction and experiment, with values above 0.991 across all $g$.

We observe that OAMNet outperforms traditional numerical simulations in matching experimental OAM spectra, especially in the high-gain regime. As shown in Fig.~\ref{fig5}k, the  MAE of OAMNet predictions drops by nearly 50\% compared to simulations for gains $g \gtrsim 2.9$. Similarly, Fig.~\ref{fig5}l shows that the cosine similarity between OAMNet and experiment consistently exceeds 0.997 in the high-gain region, surpassing simulation by up to 0.006. We attribute this to the network’s ability to learn the underlying functional dependence on the spatial mode of the pump. At a higher gain, the OAM spectrum tends to peak at the dominant mode, narrowing the width of the distribution. As a result, this makes the simulation error more sensitive to the resolution of integration meshes. Nevertheless, our PGNN treats the problem as a general regression task. By learning the fundamental dependence on the input physical parameters from the training data , it predicts the OAM distribution based on its understanding of the two-photon wavefunction on the manifold constrained by physical laws without the need of performing numerical computation. 

From a learning theory perspective, the model manages to learn and correct residual biases in the simulation model: by training on a large, smoothly varying grid of ${g,\theta,L,w_p}$ values, the PGNN acquires a calibration that compensates systematic offsets between simulated and real spectra due to the selection of integration window. In addition, the end-to-end mapping afforded by the OAMNet architecture may impose a degree of smoothness absent in discrete SVD‐based simulations, and it interpolates more flexibly across the parameter manifold. The combination of physics‐informed feature maps (OAM/radial grids) and expressive nonlinear blocks enables the network to capture secondary effects (e.g., phase‐mismatch variations) that are not fully resolved in the numerical code. Furthermore, the inclusion of experimental measurement in the training dataset can help the neural network model compensate for the numerical artifacts in the simulation data due to the limited resolution of numerical meshes when a large integration window is necessary for higher-order pump modes or when the mode distribution becomes sharply peaked at the fundamental mode. As a result, OAMNet delivers predictions that are at least as accurate as, and often closer to, experiment than the original simulation.

It is also noticeable that the OAM spectrum narrows with increasing gain. This trend has been reported in multiple experimental and theoretical studies as one of the important findings of high-gain SPDC \cite{kulkarni2022classical}.

\section{Discussion and conclusion}\label{sec5}
We have proposed and demonstrated a PGNN architecture, OAMNet, as a computationally efficient method to characterize the full modal structure of entangled photon pairs generated by high-gain SPDC in a high-dimensional Hilbert space. The proposed model is a physics-guided surrogate trained on simulation/experimental data generated from the SPDC Hamiltonian and phase-matching physics. Physical knowledge is incorporated explicitly through a soft OAM-conservation regularizer, which biases training toward physically consistent solutions without enforcing the full dynamical equations as hard architectural constraints. The conventional method to calculate the mode distribution or the OAM spectrum of a high-gain SPDC field usually involves costly operations such as high-dimensional integration and singular value decomposition of large matrices. In addition, the accuracy of traditional numerical methods also relies heavily on the proper choice of the integration window and grid coordinates. These methods can become extremely inefficient and less reliable when a large range of physical parameter space needs to be scanned. OAMNet systematically corrects residual biases in the numerical model, yields smoother spectra across both low and high OAM orders, and interpolates reliably between training points in parameter space. Compared with traditional simulation methods, our PGNN approach accelerates end‐to‐end mode‐structure prediction by more than two orders of magnitude while maintaining, or even improving, agreement with experiment. This work demonstrates that physics‐informed neural networks can surpass purely first‐principles simulations in both accuracy and computational efficiency, paving the way for real‐time design and optimization of complex quantum‐optical systems.

The field of quantum optics is now witnessing an unprecedented development in the applications of machine learning for physics. With state-of-the-art AI architectures, we expect to extend our current PGNN model to more specific and realistic experiment settings. For example, we will be able to employ OAMNet in the context of operator learning \cite{hao2023gnot, kovachki2024operator} so that the quantum nature of SPDC (e.g. quadrature squeezing and photon statistics) can be fully captured. Moreover, the study on the effect of turbulence on entanglement has significant practical implications in free-space QKD \cite{malik2012influence}. In the future, our PGNN framework can be integrated to existing turbulence models, such as RANS/LENS models \cite{5t33-dj8h}, to accurately describe the chaotic dynamics of entangled photon pairs propagating in turbulent channels. By incorporating experimental measurement, theoretical modeling, and data-driven learning, the OAMNet in this work sets up the new paradigm for future AI-aided quantum engineering.

\begin{backmatter}

\bmsection{Acknowledgment}
The authors thank Girish Kulkarni for the useful discussion on the interferometric measurement of OAM spectrum.

\bmsection{Disclosures}
The authors declare no conflicts of interest.

\bmsection{Data availability} Data underlying the results presented in this paper are not publicly available at this time but may be obtained from the authors upon reasonable request.

\bmsection{Supplemental document}
See Supplement 1 for supporting content.

\end{backmatter}


\bibliography{main}

@article{sharapova2020properties,
  title={Properties of bright squeezed vacuum at increasing brightness},
  author={Sharapova, PR and Frascella, Gaetano and Riabinin, M and P{\'e}rez, AM and Tikhonova, OV and Lemieux, S and Boyd, RW and Leuchs, G and Chekhova, MV},
  journal={Physical Review Research},
  volume={2},
  number={1},
  pages={013371},
  year={2020},
  publisher={APS}
}

@article{PhysRevA.106.063709,
  title = {Reconstructing the full modal structure of photonic states by stimulated-emission tomography in the low-gain regime},
  author = {Keller, A. and Khoury, A. Z. and Fabre, N. and Amanti, M. and Baboux, F. and Ducci, S. and Milman, P.},
  journal = {Phys. Rev. A},
  volume = {106},
  issue = {6},
  pages = {063709},
  numpages = {12},
  year = {2022},
  month = {Dec},
  publisher = {American Physical Society},
}

@article{PhysRevA.83.033816,
  title = {Full characterization of the quantum spiral bandwidth of entangled biphotons},
  author = {Miatto, Filippo M. and Yao, Alison M. and Barnett, Stephen M.},
  journal = {Phys. Rev. A},
  volume = {83},
  issue = {3},
  pages = {033816},
  numpages = {9},
  year = {2011},
  month = {Mar},
  publisher = {American Physical Society},
}

@article{even2024motion,
  title={Motion of charged particles in bright squeezed vacuum},
  author={Even Tzur, Matan and Cohen, Oren},
  journal={Light: Science \& Applications},
  volume={13},
  number={1},
  pages={41},
  year={2024},
  publisher={Nature Publishing Group UK London}
}

@article{miatto2012spatial,
  title={Spatial Schmidt modes generated in parametric down-conversion},
  author={Miatto, Filippo M and di Lorenzo Pires, H and Barnett, Stephen M and van Exter, Martin P},
  journal={The European Physical Journal D},
  volume={66},
  number={10},
  pages={263},
  year={2012},
  publisher={Springer}
}

@article{PhysRevLett.113.060503,
  title = {Free-Space Quantum Key Distribution by Rotation-Invariant Twisted Photons},
  author = {Vallone, Giuseppe and D'Ambrosio, Vincenzo and Sponselli, Anna and Slussarenko, Sergei and Marrucci, Lorenzo and Sciarrino, Fabio and Villoresi, Paolo},
  journal = {Phys. Rev. Lett.},
  volume = {113},
  issue = {6},
  pages = {060503},
  numpages = {5},
  year = {2014},
  month = {Aug},
  publisher = {American Physical Society},
}

@article{mair2001entanglement,
  title={Entanglement of the orbital angular momentum states of photons},
  author={Mair, Alois and Vaziri, Alipasha and Weihs, Gregor and Zeilinger, Anton},
  journal={Nature},
  volume={412},
  number={6844},
  pages={313--316},
  year={2001},
  publisher={Nature Publishing Group UK London}
}

@article{PhysRevLett.101.163602,
  title = {Spatial Symmetry and Conservation of Orbital Angular Momentum in Spontaneous Parametric Down-Conversion},
  author = {Feng, Sheng and Kumar, Prem},
  journal = {Phys. Rev. Lett.},
  volume = {101},
  issue = {16},
  pages = {163602},
  numpages = {4},
  year = {2008},
  month = {Oct},
  publisher = {American Physical Society},
}

@article{PhysRevLett.134.203601,
  title = {Conservation of Angular Momentum on a Single-Photon Level},
  author = {Kopf, L. and Barros, R. and Prabhakar, S. and Giese, E. and Fickler, R.},
  journal = {Phys. Rev. Lett.},
  volume = {134},
  issue = {20},
  pages = {203601},
  numpages = {7},
  year = {2025},
  month = {May},
  publisher = {American Physical Society},
}

@article{xiao1987precision,
  title={Precision measurement beyond the shot-noise limit},
  author={Xiao, Min and Wu, Ling-An and Kimble, H Jeffrey},
  journal={Physical review letters},
  volume={59},
  number={3},
  pages={278},
  year={1987},
  publisher={APS}
}

@article{oelker2016ultra,
  title={Ultra-low phase noise squeezed vacuum source for gravitational wave detectors},
  author={Oelker, E and Mansell, G and Tse, M and Miller, J and Matichard, F and Barsotti, L and Fritschel, P and McClelland, DE and Evans, M and Mavalvala, N},
  journal={Optica},
  volume={3},
  number={7},
  pages={682--685},
  year={2016},
  publisher={Optical Society of America}
}

@article{pooser2015ultrasensitive,
  title={Ultrasensitive measurement of microcantilever displacement below the shot-noise limit},
  author={Pooser, Raphael C and Lawrie, Benjamin},
  journal={Optica},
  volume={2},
  number={5},
  pages={393--399},
  year={2015},
  publisher={Optical Society of America}
}

@article{cutipa2022bright,
  title={Bright squeezed vacuum for two-photon spectroscopy: simultaneously high resolution in time and frequency, space and wavevector},
  author={Cutipa, Paula and Chekhova, Maria V},
  journal={Optics Letters},
  volume={47},
  number={3},
  pages={465--468},
  year={2022},
  publisher={Optica Publishing Group}
}

@article{heimerl2025quantum,
  title={Quantum light drives electrons strongly at metal needle tips},
  author={Heimerl, Jonas and Rasputnyi, Andrei and P{\"o}lloth, Jonathan and Meier, Stefan and Chekhova, Maria and Hommelhoff, Peter},
  journal={Nature Physics},
  pages={1--6},
  year={2025},
  publisher={Nature Publishing Group UK London}
}

@article{kulkarni2017single,
  title={Single-shot measurement of the orbital-angular-momentum spectrum of light},
  author={Kulkarni, Girish and Sahu, Rishabh and Maga{\~n}a-Loaiza, Omar S and Boyd, Robert W and Jha, Anand K},
  journal={Nature communications},
  volume={8},
  number={1},
  pages={1054},
  year={2017},
  publisher={Nature Publishing Group UK London}
}

@article{kopf2025conservation,
  title={Conservation of angular momentum on a single-photon level},
  author={Kopf, Lea and Barros, Rafael and Prabhakar, Shashi and Giese, Enno and Fickler, Robert},
  journal={Physical Review Letters},
  volume={134},
  number={20},
  pages={203601},
  year={2025},
  publisher={APS}
}

@article{sharapova2015schmidt,
  title={Schmidt modes in the angular spectrum of bright squeezed vacuum},
  author={Sharapova, P and P{\'e}rez, Angela M and Tikhonova, Olga V and Chekhova, Maria V},
  journal={Physical Review A},
  volume={91},
  number={4},
  pages={043816},
  year={2015},
  publisher={APS}
}

@article{kudyshev2020machine,
  title={Machine learning for integrated quantum photonics},
  author={Kudyshev, Zhaxylyk A and Shalaev, Vladimir M and Boltasseva, Alexandra},
  journal={ACS Photonics},
  volume={8},
  number={1},
  pages={34--46},
  year={2020},
  publisher={ACS Publications}
}

@article{zuo2022deep,
  title={Deep learning in optical metrology: a review},
  author={Zuo, Chao and Qian, Jiaming and Feng, Shijie and Yin, Wei and Li, Yixuan and Fan, Pengfei and Han, Jing and Qian, Kemao and Chen, Qian},
  journal={Light: Science \& Applications},
  volume={11},
  number={1},
  pages={39},
  year={2022},
  publisher={Nature Publishing Group UK London}
}

@article{saba2022physics,
  title={Physics-informed neural networks for diffraction tomography},
  author={Saba, Amirhossein and Gigli, Carlo and Ayoub, Ahmed B and Psaltis, Demetri},
  journal={Advanced Photonics},
  volume={4},
  number={6},
  pages={066001--066001},
  year={2022},
  publisher={Society of Photo-Optical Instrumentation Engineers}
}

@article{cong2019quantum,
  title={Quantum convolutional neural networks},
  author={Cong, Iris and Choi, Soonwon and Lukin, Mikhail D},
  journal={Nature Physics},
  volume={15},
  number={12},
  pages={1273--1278},
  year={2019},
  publisher={Nature Publishing Group UK London}
}

@article{youssry2024experimental,
  title={Experimental graybox quantum system identification and control},
  author={Youssry, Akram and Yang, Yang and Chapman, Robert J and Haylock, Ben and Lenzini, Francesco and Lobino, Mirko and Peruzzo, Alberto},
  journal={npj Quantum Information},
  volume={10},
  number={1},
  pages={9},
  year={2024},
  publisher={Nature Publishing Group UK London}
}

@article{deng2017quantum,
  title={Quantum entanglement in neural network states},
  author={Deng, Dong-Ling and Li, Xiaopeng and Das Sarma, S},
  journal={Physical Review X},
  volume={7},
  number={2},
  pages={021021},
  year={2017},
  publisher={APS}
}

@article{urena2024entanglement,
  title={Entanglement detection with classical deep neural networks},
  author={Ure{\~n}a, Julio and Sojo, Antonio and Bermejo-Vega, Juani and Manzano, Daniel},
  journal={Scientific Reports},
  volume={14},
  number={1},
  pages={18109},
  year={2024},
  publisher={Nature Publishing Group UK London}
}

@article{schmale2022efficient,
  title={Efficient quantum state tomography with convolutional neural networks},
  author={Schmale, Tobias and Reh, Moritz and G{\"a}rttner, Martin},
  journal={npj Quantum Information},
  volume={8},
  number={1},
  pages={115},
  year={2022},
  publisher={Nature Publishing Group UK London}
}

@article{quek2021adaptive,
  title={Adaptive quantum state tomography with neural networks},
  author={Quek, Yihui and Fort, Stanislav and Ng, Hui Khoon},
  journal={npj Quantum Information},
  volume={7},
  number={1},
  pages={105},
  year={2021},
  publisher={Nature Publishing Group UK London}
}

@article{salmela2021predicting,
  title={Predicting ultrafast nonlinear dynamics in fibre optics with a recurrent neural network},
  author={Salmela, Lauri and Tsipinakis, Nikolaos and Foi, Alessandro and Billet, Cyril and Dudley, John M and Genty, Go{\"e}ry},
  journal={Nature machine intelligence},
  volume={3},
  number={4},
  pages={344--354},
  year={2021},
  publisher={Nature Publishing Group UK London}
}

@article{genty2021machine,
  title={Machine learning and applications in ultrafast photonics},
  author={Genty, Go{\"e}ry and Salmela, Lauri and Dudley, John M and Brunner, Daniel and Kokhanovskiy, Alexey and Kobtsev, Sergei and Turitsyn, Sergei K},
  journal={Nature Photonics},
  volume={15},
  number={2},
  pages={91--101},
  year={2021},
  publisher={Nature Publishing Group UK London}
}

@article{boussafa2025deep,
  title={Deep learning prediction of noise-driven nonlinear instabilities in fibre optics},
  author={Boussafa, Yassin and Sader, Lynn and Hoang, Van Thuy and Chaves, Bruno P and Bougaud, Alexis and Fabert, Marc and Tonello, Alessandro and Dudley, John M and Kues, Michael and Wetzel, Benjamin},
  journal={Nature Communications},
  volume={16},
  number={1},
  pages={7800},
  year={2025},
  publisher={Nature Publishing Group UK London}
}

@article{chen2023photonic,
  title={Photonic unsupervised learning variational autoencoder for high-throughput and low-latency image transmission},
  author={Chen, Yitong and Zhou, Tiankuang and Wu, Jiamin and Qiao, Hui and Lin, Xing and Fang, Lu and Dai, Qionghai},
  journal={Science Advances},
  volume={9},
  number={7},
  pages={eadf8437},
  year={2023},
  publisher={American Association for the Advancement of Science}
}

@article{isensee2021nnu,
  title={nnU-Net: a self-configuring method for deep learning-based biomedical image segmentation},
  author={Isensee, Fabian and Jaeger, Paul F and Kohl, Simon AA and Petersen, Jens and Maier-Hein, Klaus H},
  journal={Nature methods},
  volume={18},
  number={2},
  pages={203--211},
  year={2021},
  publisher={Nature Publishing Group}
}

@inproceedings{lecun2010convolutional,
  title={Convolutional networks and applications in vision},
  author={LeCun, Yann and Kavukcuoglu, Koray and Farabet, Cl{\'e}ment},
  booktitle={Proceedings of 2010 IEEE International Symposium on Circuits and Systems},
  pages={253--256},
  year={2010},
  organization={IEEE}
}

@article{chen2020physics,
  title={Physics-informed neural networks for inverse problems in nano-optics and metamaterials},
  author={Chen, Yuyao and Lu, Lu and Karniadakis, George Em and Dal Negro, Luca},
  journal={Optics express},
  volume={28},
  number={8},
  pages={11618--11633},
  year={2020},
  publisher={OSA}
}

@article{zhang2025multi,
  title={Multi-modality deep learning for pulse prediction in homogeneous nonlinear systems via parametric conversion},
  author={Zhang, Hao and Sun, Linshan and Hirschman, Jack and Carbajo, Sergio},
  journal={APL Photonics},
  volume={10},
  number={5},
  year={2025},
  publisher={AIP Publishing}
}

@article{zhang2025hybrid,
  title={Hybrid Deep Reconstruction for Vignetting-Free Upconversion Imaging through Scattering in ENZ Materials},
  author={Zhang, Hao and Xu, Yang and Zhang, Wenwen and Choudhary, Saumya and Alam, M Zahirul and Nguyen, Long D and Klein, Matthew and Vangala, Shivashankar and Miller, J Keith and Johnson, Eric G and others},
  journal={arXiv preprint arXiv:2508.13096},
  year={2025}
}

@article{tang2022physics,
  title={Physics-informed recurrent neural network for time dynamics in optical resonances},
  author={Tang, Yingheng and Fan, Jichao and Li, Xinwei and Ma, Jianzhu and Qi, Minghao and Yu, Cunxi and Gao, Weilu},
  journal={Nature computational science},
  volume={2},
  number={3},
  pages={169--178},
  year={2022},
  publisher={Nature Publishing Group US New York}
}

@article{ji2023recent,
  title={Recent advances in metasurface design and quantum optics applications with machine learning, physics-informed neural networks, and topology optimization methods},
  author={Ji, Wenye and Chang, Jin and Xu, He-Xiu and Gao, Jian Rong and Gr{\"o}blacher, Simon and Urbach, H Paul and Adam, Aur{\`e}le JL},
  journal={Light: Science \& Applications},
  volume={12},
  number={1},
  pages={169},
  year={2023},
  publisher={Nature Publishing Group UK London}
}

@article{yuan2023geometric,
  title={Geometric deep optical sensing},
  author={Yuan, Shaofan and Ma, Chao and Fetaya, Ethan and Mueller, Thomas and Naveh, Doron and Zhang, Fan and Xia, Fengnian},
  journal={Science},
  volume={379},
  number={6637},
  pages={eade1220},
  year={2023},
  publisher={American Association for the Advancement of Science}
}

@article{zhang2025accurate,
  title={Accurate label free classification of cancerous extracellular vesicles using nanoaperture optical tweezers and deep learning},
  author={Zhang, Hao and Zhao, Tianyu and Zhang, Wenwen and Halvaei, Sina and Peters, Matthew and Cheung, Tsz Shing and Williams, Karla C and Gordon, Reuven},
  journal={npj Biosensing},
  volume={2},
  number={1},
  pages={33},
  year={2025},
  publisher={Nature Publishing Group UK London}
}

@article{lin2018all,
  title={All-optical machine learning using diffractive deep neural networks},
  author={Lin, Xing and Rivenson, Yair and Yardimci, Nezih T and Veli, Muhammed and Luo, Yi and Jarrahi, Mona and Ozcan, Aydogan},
  journal={Science},
  volume={361},
  number={6406},
  pages={1004--1008},
  year={2018},
  publisher={American Association for the Advancement of Science}
}

@article{momeni2025training,
  title={Training of physical neural networks},
  author={Momeni, Ali and Rahmani, Babak and Scellier, Benjamin and Wright, Logan G and McMahon, Peter L and Wanjura, Clara C and Li, Yuhang and Skalli, Anas and Berloff, Natalia G and Onodera, Tatsuhiro and others},
  journal={Nature},
  volume={645},
  number={8079},
  pages={53--61},
  year={2025},
  publisher={Nature Publishing Group UK London}
}

@article{barbastathis2019use,
  title={On the use of deep learning for computational imaging},
  author={Barbastathis, George and Ozcan, Aydogan and Situ, Guohai},
  journal={Optica},
  volume={6},
  number={8},
  pages={921--943},
  year={2019},
  publisher={Optical Society of America}
}

@article{brida2010experimental,
  title={Experimental realization of sub-shot-noise quantum imaging},
  author={Brida, Giorgio and Genovese, Marco and Ruo Berchera, Ivano},
  journal={Nature Photonics},
  volume={4},
  number={4},
  pages={227--230},
  year={2010},
  publisher={Nature Publishing Group UK London}
}

@article{rasputnyi2024high,
  title={High-harmonic generation by a bright squeezed vacuum},
  author={Rasputnyi, Andrei and Chen, Zhaopin and Birk, Michael and Cohen, Oren and Kaminer, Ido and Kr{\"u}ger, Michael and Seletskiy, Denis and Chekhova, Maria and Tani, Francesco},
  journal={Nature Physics},
  volume={20},
  number={12},
  pages={1960--1965},
  year={2024},
  publisher={Nature Publishing Group UK London}
}

@article{lemieux2025photon,
  title={Photon bunching in high-harmonic emission controlled by quantum light},
  author={Lemieux, Samuel and Jalil, Sohail A and Purschke, David N and Boroumand, Neda and Hammond, TJ and Villeneuve, David and Naumov, Andrei and Brabec, Thomas and Vampa, Giulio},
  journal={Nature Photonics},
  pages={1--5},
  year={2025},
  publisher={Nature Publishing Group UK London}
}

@article{lyu2025effect,
  title={Effect of photon quantum statistics on electrons in above-threshold ionization},
  author={Lyu, Zijian and Sun, Fengxiao and Fang, Yiqi and He, Qiongyi and Liu, Yunquan},
  journal={Physical Review Research},
  volume={7},
  number={1},
  pages={L012072},
  year={2025},
  publisher={APS}
}

@article{lopaeva2013experimental,
  title={Experimental realization of quantum illumination},
  author={Lopaeva, ED and Ruo Berchera, Ivano and Degiovanni, Ivo Pietro and Olivares, S and Brida, Giorgio and Genovese, Marco},
  journal={Physical Review Letters},
  volume={110},
  number={15},
  pages={153603},
  year={2013},
  publisher={APS}
}

@article{brida2010detection,
  title={Detection of multimode spatial correlation in PDC and application to the absolute calibration of a CCD camera},
  author={Brida, Giorgio and Degiovanni, Ivo Pietro and Genovese, Marco and Rastello, Maria Luisa and Ruo-Berchera, Ivano},
  journal={Optics Express},
  volume={18},
  number={20},
  pages={20572--20584},
  year={2010},
  publisher={Optical Society of America}
}

@article{PhysRevA.47.R2472,
  title = {High-visibility interference in a Bell-inequality experiment for energy and time},
  author = {Kwiat, P. G. and Steinberg, A. M. and Chiao, R. Y.},
  journal = {Phys. Rev. A},
  volume = {47},
  issue = {4},
  pages = {R2472--R2475},
  numpages = {0},
  year = {1993},
  month = {Apr},
  publisher = {American Physical Society},
}

@article{PhysRevA.54.R1,
  title = {Postselection-free energy-time entanglement},
  author = {Strekalov, D. V. and Pittman, T. B. and Sergienko, A. V. and Shih, Y. H. and Kwiat, P. G.},
  journal = {Phys. Rev. A},
  volume = {54},
  issue = {1},
  pages = {R1--R4},
  numpages = {0},
  year = {1996},
  month = {Jul},
  publisher = {American Physical Society},
}

@article{PhysRevLett.92.127903,
  title = {Analysis and Interpretation of High Transverse Entanglement in Optical Parametric Down Conversion},
  author = {Law, C. K. and Eberly, J. H.},
  journal = {Phys. Rev. Lett.},
  volume = {92},
  issue = {12},
  pages = {127903},
  numpages = {4},
  year = {2004},
  month = {Mar},
  publisher = {American Physical Society},

}

@article{PhysRevA.76.062305,
  title = {Transverse spatial entanglement in parametric down-conversion},
  author = {Walborn, S. P. and Monken, C. H.},
  journal = {Phys. Rev. A},
  volume = {76},
  issue = {6},
  pages = {062305},
  numpages = {6},
  year = {2007},
  month = {Dec},
  publisher = {American Physical Society},
}

@article{PhysRevA.50.5122,
  title = {Theory of two-photon entanglement in type-II optical parametric down-conversion},
  author = {Rubin, Morton H. and Klyshko, David N. and Shih, Y. H. and Sergienko, A. V.},
  journal = {Phys. Rev. A},
  volume = {50},
  issue = {6},
  pages = {5122--5133},
  numpages = {0},
  year = {1994},
  month = {Dec},
  publisher = {American Physical Society},
}

@article{PhysRevA.69.041801,
  title = {Generation of ultrabright tunable polarization entanglement without spatial, spectral, or temporal constraints},
  author = {Fiorentino, Marco and Messin, Ga\'etan and Kuklewicz, Christopher E. and Wong, Franco N. C. and Shapiro, Jeffrey H.},
  journal = {Phys. Rev. A},
  volume = {69},
  issue = {4},
  pages = {041801},
  numpages = {4},
  year = {2004},
  month = {Apr},
  publisher = {American Physical Society},
}

@article{fickler2012quantum,
  title={Quantum entanglement of high angular momenta},
  author={Fickler, Robert and Lapkiewicz, Radek and Plick, William N and Krenn, Mario and Schaeff, Christoph and Ramelow, Sven and Zeilinger, Anton},
  journal={Science},
  volume={338},
  number={6107},
  pages={640--643},
  year={2012},
  publisher={American Association for the Advancement of Science}
}

@article{romero2012increasing,
  title={Increasing the dimension in high-dimensional two-photon orbital angular momentum entanglement},
  author={Romero, J and Giovannini, D and Franke-Arnold, S and Barnett, SM and Padgett, MJ},
  journal={Physical Review A},
  volume={86},
  number={1},
  pages={012334},
  year={2012},
  publisher={APS}
}

@ARTICLE{spec,
  title={Full characterization of the quantum spiral bandwidth of entangled biphotons},
  author={Miatto, Filippo M and Yao, Alison M and Barnett, Stephen M},
  journal={Physical Review A},
  volume={83},
  number={3},
  pages={033816},
  year={2011},
  publisher={APS}
}

@article{PhysRevLett.102.183602,
  title = {Generation and Direct Detection of Broadband Mesoscopic Polarization-Squeezed Vacuum},
  author = {Iskhakov, Timur and Chekhova, Maria V. and Leuchs, Gerd},
  journal = {Phys. Rev. Lett.},
  volume = {102},
  issue = {18},
  pages = {183602},
  numpages = {4},
  year = {2009},
  month = {May},
  publisher = {American Physical Society},
}

@article{kulkarni2022classical,
  title={Classical model of spontaneous parametric down-conversion},
  author={Kulkarni, Girish and Rioux, Jeremy and Braverman, Boris and Chekhova, Maria V and Boyd, Robert W},
  journal={Physical Review Research},
  volume={4},
  number={3},
  pages={033098},
  year={2022},
  publisher={APS}
}

@article{PhysRevA.82.011801,
  title = {Two-color bright squeezed vacuum},
  author = {Agafonov, Ivan N. and Chekhova, Maria V. and Leuchs, Gerd},
  journal = {Phys. Rev. A},
  volume = {82},
  issue = {1},
  pages = {011801},
  numpages = {4},
  year = {2010},
  month = {Jul},
  publisher = {American Physical Society},
}

@article{hurvitz2023frequency,
  title={Frequency-domain engineering of bright squeezed vacuum for continuous-variable quantum information},
  author={Hurvitz, Inbar and Karnieli, Aviv and Arie, Ady},
  journal={Optics Express},
  volume={31},
  number={12},
  pages={20387--20397},
  year={2023},
  publisher={Optica Publishing Group}
}

@article{lawrie2019quantum,
  title={Quantum sensing with squeezed light},
  author={Lawrie, Benjamin J and Lett, Paul D and Marino, Alberto M and Pooser, Raphael C},
  journal={Acs Photonics},
  volume={6},
  number={6},
  pages={1307--1318},
  year={2019},
  publisher={ACS Publications}
}

@article{kovachki2024operator,
  title={Operator learning: Algorithms and analysis},
  author={Kovachki, Nikola B and Lanthaler, Samuel and Stuart, Andrew M},
  journal={Handbook of Numerical Analysis},
  volume={25},
  pages={419--467},
  year={2024},
  publisher={Elsevier}
}

@inproceedings{hao2023gnot,
  title={Gnot: A general neural operator transformer for operator learning},
  author={Hao, Zhongkai and Wang, Zhengyi and Su, Hang and Ying, Chengyang and Dong, Yinpeng and Liu, Songming and Cheng, Ze and Song, Jian and Zhu, Jun},
  booktitle={International Conference on Machine Learning},
  pages={12556--12569},
  year={2023},
  organization={PMLR}
}

@article{malik2012influence,
  title={Influence of atmospheric turbulence on optical communications using orbital angular momentum for encoding},
  author={Malik, Mehul and O’Sullivan, Malcolm and Rodenburg, Brandon and Mirhosseini, Mohammad and Leach, Jonathan and Lavery, Martin PJ and Padgett, Miles J and Boyd, Robert W},
  journal={Optics express},
  volume={20},
  number={12},
  pages={13195--13200},
  year={2012},
  publisher={Optical Society of America}
}

@article{5t33-dj8h,
  title = {${\ensuremath{\ell}}^{2}/{g}^{2}$ hybrid {RANS}/{LES} model for simulating turbulent flows in the spectral element framework},
  author = {Wang, Sijie and Cheng, Yuxiao and Yin, Zifei and Durbin, Paul and Li, Weipeng},
  journal = {Phys. Rev. Fluids},
  volume = {10},
  issue = {9},
  pages = {094902},
  numpages = {23},
  year = {2025},
  month = {Sep},
  publisher = {American Physical Society},

}

\end{document}